\documentclass[preprint,preprintnumbers,amsmath,amssymb,nofootinbib]{revtex4}

\usepackage{graphicx}
\usepackage{amsmath}
\usepackage{amssymb}
\usepackage{graphicx}
\usepackage{dsfont}
\usepackage{overpic}
\usepackage{color}
\usepackage{slashed}
\usepackage{subfigure}
\usepackage{graphicx,textcomp}
\usepackage{natbib}
\usepackage{extarrows}

\allowdisplaybreaks

\newcommand{\newop}[2]{\def#1{\mathop{\mathrm{#2}}\nolimits}}
\newop{\artanh}{artanh}
\newop{\det}{det}

\begin{document}
\title{A simple model for a scalar two-point correlator in the presence of a resonance}
\author{Peter C.~Bruns}
\affiliation{Institut f\"ur Theoretische Physik, Universit\"at Regensburg, D-93040 Regensburg, Germany}
\date{\today}
\begin{abstract}
We present a simple toy model for a scalar-isoscalar two-point correlator, which can serve as a testing ground for the extraction of resonance parameters from Lattice QCD calculations. We discuss in detail how the model correlator behaves when it is restricted to a finite spatial volume, and how the finite-volume data can be used to reconstruct the spectral function of the correlator in the infinite volume, which allows to extract properties of the resonance from such data.
\end{abstract}
\maketitle

\section{Introduction and disclaimer}

In the past years, the extraction of information on hadronic resonances from Lattice QCD data has become of great interest. Recent lattice studies of this subject can e.g. be found in \cite{Feng:2010es,Lang:2011mn,Gockeler:2012yj,Mohler:2012nh,Dudek:2012xn,Lang:2015sba,Wilson:2015dqa,Bali:2015gji,Guo:2016zos,Briceno:2016mjc,Wu:2016ixr,Braun:2016wnx}, and references therein. In Quantum Field Theory (QFT), resonances are associated with poles of scattering or transition amplitudes on unphysical Riemann sheets of the complex energy surface pertaining to these amplitudes, and all properties of the resonance should refer somehow to these poles in the complex plane \cite{Gegelia:2009py}. In a finite volume, however, the energy spectrum is discrete, and the correlators and amplitudes are purely real in euclidean time. There are no branch cuts and no Riemann sheets, and it therefore seems nontrivial to relate the measurements in a finite box to the resonance phenomena observed in (or extracted from) experimental data. 
Of course, the theory needed to close this gap is already well-developed, see \cite{Maiani:1990ca,Luscher:1990ux,Luscher:1991cf,Wiese:1988qy,Lellouch:2000pv} for the ``classic'' articles on the subject, and \cite{Meissner:2010rq,Bernard:2010fp,Doring:2011vk,Chen:2012rp,Hansen:2012tf,Bernard:2012bi,Briceno:2014uqa,Bolton:2015psa,Briceno:2015tza,Agadjanov:2016mao,Guo:2016zep,Wang:2017tep} for some recent developments. In the literature just cited, the problem with which we deal here is examined in much more depth and generality than is attempted in the present article. Here, we focus on a specific simple field-theoretic toy model - perhaps the simplest one that shows all the features we want to study on a basic level (resonance dynamics in a finite volume, extraction of data related to a form factor in the time-like region in the presence of a resonance), but is still consistent with the field-theoretical strictures of unitarity and analyticity. Our first aim is to provide the practitioner (i.e., people who are concerned with the analysis of realistic lattice data) with an explicit model, as a testing ground for the methods used to deal with real hadronic resonances in Lattice QCD. Our second aim is to present an amenable representation of a non-trivial (momentum-projected) two-point correlator which can be used for pedagogic purposes, since we have often noticed that the relation between resonance physics and finite-volume, euclidean-time lattice data is often hard to understand for students in the field of Lattice QCD, or even for researchers in related areas. Our third aim is to make some technical details of the relevant derivations and calculations available, which are often hard to retrace from the original papers on the subject. Here we try to be as explicit as possible. 
We do not treat discretization effects, and we shall neglect, for the most part, those finite-volume effects which are exponentially suppressed with the spatial extent of the finite box, like e.g. finite-volume corrections to masses and coupling constants. The methods to deal with the latter modifications are very well-known, see e.g. \cite{Luscher:1985dn,Hasenfratz:1989pk,Colangelo:2005gd}.\\
 \\
The plan of this article is as follows: In Sec.~\ref{sec:generalities}, we introduce the general framework and the basic properties of the two-point correlator in our model. In Sec.~\ref{sec:T}, we outline the construction of the amplitude used to describe the two-body scattering between the elementary, stable particles of our theory (called $\phi$ particles here). The model for the correlator and its behavior in a finite volume are discussed in detail in Sec.~\ref{sec:M}. In Sec.~\ref{sec:Num}, we go through a numerical demonstration of the methods and results obtained in the previous sections, and give a short conclusion and outlook.

\newpage

\section{General properties of the scalar two-point correlator}
\label{sec:generalities}
We consider the two-point correlator of a scalar operator $\mathcal{S}(x)$, 
\begin{equation}\label{eq:c}
c_{SS}(x-y):=\langle 0|T\mathcal{S}(x)\mathcal{S}(y)|0\rangle\,,
\end{equation}
within a model field theory of pseudoscalar $\phi$ particles of mass $M$. We assume that there is an s-wave resonance, called $\sigma$ here, which occurs in $\phi\phi\rightarrow\phi\phi$ scattering. For simplicity, the only asymptotic states for which the operator $\mathcal{S}(x)$ has non-vanishing matrix elements with the vacuum state are assumed to be the two-particle $\phi\phi$ states, $\langle 0|\mathcal{S}(x)|\phi(\mathbf{k}_{1})\phi(\mathbf{k}_{2})\rangle\not=0$ (however, in the case where the $\sigma$ becomes a bound state instead of a resonance, we shall also admit a $\sigma$ state and $\langle 0|\mathcal{S}(x)|\sigma(\mathbf{k})\rangle\not=0$). We are, in particular, interested in the value of the matrix elements for a specific prescribed three-momentum $\mathbf{p}$. From now on we let $y=(0,\mathbf{0})$, and denote the ``momentum-projected'' matrix element as
\begin{equation}\label{eq:cp}
\tilde{c}_{SS}(t,\mathbf{p}) := \int d^{3}\mathbf{x}\,e^{-i\mathbf{p}\cdot\mathbf{x}}c_{SS}(t,\mathbf{x})\,.
\end{equation}
According to the general rules of Quantum Field Theory\footnote{See e.g. the textbook \cite{Weinberg:1995mt}, in particular Chapter 6 and Eqs.~(10.4.19,20) therein, and also Chapter 9 of \cite{Itzykson:1980rh}.}, the correlator can be represented as
\begin{equation}\label{eq:cM}
c_{SS}(t,\mathbf{x}) = \int\frac{d^{4}q}{(2\pi)^4}e^{-iqx}i\mathcal{M}(q^2)\,,
\end{equation}
where $i\mathcal{M}(q)$ is the sum of all Feynman graphs (in the given field theory) with operator insertions at two fixed space-time positions $x=(t,\mathbf{x})$, $t>0$, and $y=(0,\mathbf{0})$, and where $q=(q^{0},\mathbf{q})$ can be interpreted as the four-momentum flowing into the operator insertion at $y$ and out of the operator insertion at $x$. The vertex rule for the operator insertion coupling to $\phi\phi$ is simply given by a constant $b$ in our model (and $\tilde{b}$ in the case of the $\sigma$), so that {\em in the limit where the interactions are ``turned off''\,}, we have just $\langle 0|\mathcal{S}(0)|\phi(\mathbf{k}_{1})\phi(\mathbf{k}_{2})\rangle\rightarrow b$ and $\langle 0|\mathcal{S}(0)|\sigma(\mathbf{k})\rangle\rightarrow\tilde{b}$. \\
For example, the exchange of an ``undressed'' $\sigma$ particle (with mass $m_{0}$) gives a contribution 
\begin{equation}\label{eq:baresigmaex}
i\mathcal{M}_{\sigma-\mathrm{ex.}} = \frac{i\tilde{b}^2}{q^2-m_{0}^{2}+i\epsilon}\quad\Rightarrow\quad \tilde{c}_{SS}(t,\mathbf{p})|_{\sigma-\mathrm{ex.}} = \frac{\tilde{b}^2 e^{-it\sqrt{|\mathbf{p}|^2+m_{0}^{2}}}}{2\sqrt{|\mathbf{p}|^2+m_{0}^{2}}}\,,
\end{equation}
where we have used the appropriate $i\epsilon$ prescription. The relevant Fourier integral can be inferred from Eq.~(\ref{eq:cosintz}).
\newpage
\subsection*{Exchange of a pair of non-interacting $\phi$ particles}
From the exchange of a pair of free pseudoscalar $\phi$ particles between the two vertex insertions results the contribution
\begin{eqnarray}
i\mathcal{M}_{\phi\phi-\mathrm{ex.}} &=& i\frac{b^2}{2}\int\frac{d^{d}l}{(2\pi)^{d}}\frac{i}{[(q-l)^2-M^2+i\epsilon][l^2-M^2+i\epsilon]}\nonumber \\
 &\overset{d\rightarrow 4}{=}& ib^2\left(I_{\phi\phi}(q^2=4M^2) + \frac{4M^2-q^2}{32\pi^2}\int_{4M^2}^{\infty}ds'\,\frac{\sigma(s')}{(s'-4M^2)(s'-(q^2+i\epsilon))}\right) \nonumber \\
 &=:& ib^2\left(I_{\phi\phi}(q^2=4M^2) + \bar{I}_{\phi\phi}(q^2)\right)\,,\qquad \sigma(s') := \sqrt{1-\frac{4M^2}{s'}}\,.\label{eq:fifi1}
\end{eqnarray}
The first term in the second line, $I_{\phi\phi}(q^2=4M^2)$, contains a divergent constant for $d\rightarrow 4$, which results in a contact term contribution $\sim\delta(t)$ in $\tilde{c}_{SS}(t,\mathbf{p})$. As we are interested in the behavior of the correlator for large positive times, this constant term can be dropped. Then,
\begin{eqnarray}
\tilde{c}_{SS}(t,\mathbf{p})|_{\phi\phi-\mathrm{ex.}} &\overset{t>0}{=}& i\frac{b^2}{2}\int_{4M^2}^{\infty}ds'\,\frac{\sigma(s')}{32\pi^3(s'-4M^2)}\int_{-\infty}^{+\infty}dq^{0}\,\frac{((q^{0})^2-(|\mathbf{p}|^2+4M^2))\,e^{-iq^{0}t}}{(q^{0})^2+i\epsilon -(|\mathbf{p}|^2+s')} \nonumber \\
 &=& \frac{b^2}{\pi}\int_{4M^2}^{\infty}ds'\,\left(\frac{\sigma(s')}{32\pi}\right)\frac{e^{-i\sqrt{s'+|\mathbf{p}|^2}t}}{2\sqrt{s'+|\mathbf{p}|^2}} \nonumber \\
 &=& \frac{b^2}{32\pi^2}\int_{\varepsilon_{p}^{\phi\phi}}^{\infty}dE'\,\frac{\sqrt{E'^2-(|\mathbf{p}|^2+4M^2)}}{\sqrt{E'^2-|\mathbf{p}|^2}}\,e^{-iE't}\,.\label{eq:fifi2}
\end{eqnarray}
In the last line, we have introduced $\varepsilon_{p}^{\phi\phi}:=\sqrt{|\mathbf{p}|^2+4M^{2}}\,$ and a new integration variable $E':=\sqrt{s'+|\mathbf{p}|^2}$. Integrals of the type of this result are studied in App.~\ref{app:moreints}. From those results, we can infer the behavior of our matrix element contribution for large positive times $t$ (see Eqs.~(\ref{eq:gasymp}) and~(\ref{eq:gtildeasymp})-(\ref{eq:expdintasymp}))
\begin{equation}\label{eq:cploop}
\tilde{c}_{SS}(t,\mathbf{p})\bigr|_{\phi\phi-\mathrm{ex.}}\quad\overset{t\rightarrow\infty}{\longrightarrow}\quad \frac{b^2}{64\pi^{\frac{3}{2}}}\frac{\sqrt{2\varepsilon_{p}^{\phi\phi}}}{2M}\frac{e^{-i\varepsilon_{p}^{\phi\phi}t}}{(it)^{\frac{3}{2}}}\,.
\end{equation}
It is reassuring to see that the result is real and positive for euclidean times $\tau=it$.\\
Given that the full amplitude satisfies a dispersive representation
\begin{equation}\label{eq:dispM}
\mathrm{Im}\,\mathcal{M}(q^2)=-B^{\ast}(q^2)\frac{\sigma(q^2)}{32\pi}B(q^2)\,,\qquad \mathcal{M}(q^2)=\frac{1}{\pi}\int_{4M^2}^{\infty}ds'\,\frac{\mathrm{Im}\,\mathcal{M}(s')}{s'-(q^2+i\epsilon)}\,,
\end{equation}
with some complex function $B(q^2)$, the above result can be readily generalized,
\begin{equation}\label{eq:cfull}
\tilde{c}_{SS}(t,\mathbf{p}) = \frac{1}{32\pi^2}\int_{\varepsilon_{p}^{\phi\phi}}^{\infty}dE'\,\frac{\sqrt{E'^2-(|\mathbf{p}|^2+4M^2)}}{\sqrt{E'^2-|\mathbf{p}|^2}}\,|B(E')|^2\,e^{-iE't}\,.
\end{equation}
Again, given Eq.~(\ref{eq:dispM}), we find that $\tilde{c}_{SS}(t,\mathbf{p})$ is real and positive for euclidean times $\tau=it$.
\newpage
\subsection*{Exchange of a pair of non-interacting $\phi$ particles - in a finite volume}
In a cubic box with side length $L$, employing periodic boundary conditions, the three-momenta are quantized, so that e.g. $\mathbf{q}=\frac{2\pi}{L}\mathbf{N}$, where $\mathbf{N}\in\mathds{Z}^{3}$. Accordingly, the momentum integrals now correspond to discrete summations, $\int d^{3}l\,\rightarrow\,\left(\frac{2\pi}{L}\right)^3\sum_{\mathbf{n}\in\mathds{Z}^3}$ with quantized loop momenta. Consider the function
\begin{eqnarray}
I_{\phi\phi}^{L}(x;q) &:=& \frac{1}{2}\left(\frac{2\pi}{L}\right)^3\int\frac{d^4l}{(2\pi)^{4}}\frac{ie^{-il x}\sum_{\mathbf{n}\in\mathds{Z}^3}\delta^{3}\left(\mathbf{l}-\frac{2\pi}{L}\left(\mathbf{n}+\frac{1}{2}\mathbf{N}\right)\right)}{\left[\left(\frac{q}{2}-l\right)^2-M^2+i\epsilon\right]\left[\left(\frac{q}{2}+l\right)^2-M^2+i\epsilon\right]}\nonumber \\
 &=& \frac{1}{2}\int\frac{d^4l}{(2\pi)^{4}}\frac{ie^{-ilx}\sum_{\mathbf{k}\in\mathds{Z}^3}e^{i\mathbf{k}\cdot\left(L\mathbf{l}-\pi\mathbf{N}\right)}}{\left[\left(\frac{q}{2}-l\right)^2-M^2+i\epsilon\right]\left[\left(\frac{q}{2}+l\right)^2-M^2+i\epsilon\right]} \label{eq:immqfv}\\
 &=& -\frac{1}{16\pi^2}\sum_{\mathbf{k}\in\mathds{Z}^3}\int_{-\frac{1}{2}}^{+\frac{1}{2}}dz\,e^{-iqxz}e^{i\pi\mathbf{k}\cdot\mathbf{N}(2z-1)}K_{0}\left(M(z)\sqrt{|\mathbf{x}+L\mathbf{k}|^2-t^2}\right)\,,\nonumber
\end{eqnarray}
where $x=(t,\mathbf{x})$ and $M(z)= \sqrt{M^2-\frac{q^2}{4}+q^2z^2-i\epsilon}$. We have used the Poisson summation formula in the second line, and introduced a Feynman parameter $z$ in the third line. The modified Bessel function $K_{0}$ appears due to the results of App.~\ref{app:fourier}. For $q^2>4M^2$, $M(z)$ is not real in the whole integration range, and the summations do not necessarily converge there. Let us limit ourselves to $q^2\leq 4M^2$ for a moment, and let $t\rightarrow 0$. For $L\rightarrow\infty$, only the term with $\mathbf{k}=\mathbf{0}$ survives, and so we find
\begin{eqnarray}
I_{\phi\phi}^{\infty}(x;0) &=& -\frac{1}{16\pi^2}\int_{-\frac{1}{2}}^{+\frac{1}{2}}dz\,K_{0}\left(M|\mathbf{x}|\right) = -\frac{K_{0}\left(M|\mathbf{x}|\right)}{16\pi^2} = \frac{1}{16\pi^2}\left(\log\left(\frac{M|\mathbf{x}|}{2}\right)+\gamma_{E}\right)+\mathcal{O}(|\mathbf{x}|^2)\,,\nonumber \\
I_{\phi\phi}^{\infty}(x;0) &-& I_{\phi\phi}^{L}(x;0) = \frac{1}{16\pi^2}\sum_{\mathbf{0}\not=\mathbf{k}\in\mathds{Z}^3}K_{0}\left(M|\mathbf{x}+L\mathbf{k}|\right)\,,\nonumber 
\end{eqnarray}
while for $q\rightarrow (2M,\mathbf{0})$, we find
\begin{equation*}
I_{\phi\phi}^{\infty}(x;(2M,\mathbf{0})) = -\frac{1}{16\pi^2}\int_{-\frac{1}{2}}^{+\frac{1}{2}}dz\,K_{0}\left(2M|z||\mathbf{x}|\right) = \frac{1}{16\pi^2}\left(\log\left(\frac{M|\mathbf{x}|}{2}\right)+\gamma_{E} -1\right)+\mathcal{O}(|\mathbf{x}|^2)\,,
\end{equation*}
so that in the difference $\mathrm{lim}_{x\rightarrow 0}\left(I_{\phi\phi}^{\infty}(x;0)-I_{\phi\phi}^{\infty}(x;(2M,\mathbf{0}))\right)=(16\pi^2)^{-1}$, the regulator $|\mathbf{x}|^{-1}$ drops out. Starting from the first line of Eq.~(\ref{eq:immqfv}), we use the residue theorem to compute
\begin{eqnarray}
I_{\phi\phi}^{L}(x;q) &-& I_{\phi\phi}^{L}(x;0) \,\overset{x\rightarrow 0}{\longrightarrow}\, \frac{1}{2L^3}\sum_{\mathbf{n}\in\mathds{Z}^3}\left(\frac{E_{+}^{(\mathbf{n})}+E_{-}^{(\mathbf{n})}}{2E_{+}^{(\mathbf{n})}E_{-}^{(\mathbf{n})}(q^2+i\epsilon -((E_{+}^{(\mathbf{n})}+E_{-}^{(\mathbf{n})})^2-|\mathbf{q}|^2)}+\frac{1}{4(E_{0}^{(\mathbf{n})})^3}\right)\,,   \nonumber \\
E_{\pm}^{(\mathbf{n})} &=& \sqrt{\left(\frac{2\pi}{L}\right)^2\left|\frac{\mathbf{N}}{2}\pm\left(\mathbf{n}+\frac{\mathbf{N}}{2}\right)\right|^2+M^2}\,,\quad E_{0}^{(\mathbf{n})} = \sqrt{\left|\frac{2\pi}{L}\mathbf{n}\right|^2+M^2}\,,
\end{eqnarray}
and where $q^2\equiv(q^{0})^2-|\mathbf{q}|^2$, $\mathbf{q}=\frac{2\pi}{L}\mathbf{N}$\,. Combining the above observations, we are in a position to write down the finite-volume generalization $\bar{I}_{\phi\phi}^{fv}(q,L)$ of the subtracted (infinite-volume) loop function $\bar{I}_{\phi\phi}(q^2)$:
\begin{eqnarray}
\bar{I}_{\phi\phi}^{fv}(q,L) &:=& I_{\phi\phi}^{L}(0;q) - I_{\phi\phi}^{L}(0;0) + \left(I_{\phi\phi}^{L}(0;0)-I_{\phi\phi}^{\infty}(0;0)\right) + \left(I_{\phi\phi}^{\infty}(0;0)-I_{\phi\phi}^{\infty}(0;(2M,\mathbf{0})) \right) \nonumber \\
 &=& \frac{1}{16\pi^2}\left(1-\sum_{\mathbf{0}\not=\mathbf{k}\in\mathds{Z}^3}K_{0}\left(ML|\mathbf{k}|\right)\right) \nonumber \\ &+& \frac{1}{L^3}\sum_{\mathbf{n}\in\mathds{Z}^3}\left(\frac{E_{+}^{(\mathbf{n})}+E_{-}^{(\mathbf{n})}}{4E_{+}^{(\mathbf{n})}E_{-}^{(\mathbf{n})}(q^2+i\epsilon -((E_{+}^{(\mathbf{n})}+E_{-}^{(\mathbf{n})})^2-|\mathbf{q}|^2)}+\frac{1}{(2E_{0}^{(\mathbf{n})})^3}\right) \,.\label{eq:ifififv1}
\end{eqnarray}
In this way, we have assured that the subtraction $I_{\phi\phi}^{\infty}(0;(2M,\mathbf{0}))=I_{\phi\phi}(q^2=4M^2)$ is the same in finite and in infinite volume, so that the ``renormalization'' does not depend on the volume $L^3$. Keeping $|\mathbf{N}|\sim|\mathbf{q}|$ fixed, we can write this in a dispersive form as in Eqs.~(\ref{eq:fifi1}) and (\ref{eq:fifi2}):
\begin{eqnarray}
\bar{I}_{\phi\phi}^{fv}(q,L) &=& \iota_{\mathbf{q}} - \frac{q^2}{\pi}\int_{4M^2}^{\infty}ds'\,\frac{\sum_{\mathbf{n}\in\mathds{Z}^3}\left(\frac{\pi\left(E_{+}^{(\mathbf{n})}+E_{-}^{(\mathbf{n})}\right)}{4L^3E_{+}^{(\mathbf{n})}E_{-}^{(\mathbf{n})}}\right)\delta\left(s'-s_{\mathbf{n}}^{(0)}(\mathbf{q})\right)}{s'(s'-(q^2+i\epsilon))}\,,\nonumber \\
 \iota_{\mathbf{q}} &=& \frac{1}{16\pi^2}\left(1-\sum_{\mathbf{0}\not=\mathbf{k}\in\mathds{Z}^3}K_{0}\left(ML|\mathbf{k}|\right)\right) + \sum_{\mathbf{n}\in\mathds{Z}^3}\frac{1}{(2E_{0}^{(\mathbf{n})}L)^3} \nonumber \\ &-& \frac{1}{\pi}\int_{4M^2}^{\infty}\frac{ds'}{s'}\sum_{\mathbf{n}\in\mathds{Z}^3}\left(\frac{\pi\left(E_{+}^{(\mathbf{n})}+E_{-}^{(\mathbf{n})}\right)}{4L^3E_{+}^{(\mathbf{n})}E_{-}^{(\mathbf{n})}}\right)\delta\left(s'-s_{\mathbf{n}}^{(0)}(\mathbf{q})\right)\,,\nonumber \\
 s_{\mathbf{n}}^{(0)}(\mathbf{q}) &:=& (E_{+}^{(\mathbf{n})}+E_{-}^{(\mathbf{n})})^2-|\mathbf{q}|^2\,.\label{eq:ifififv2}
\end{eqnarray}
The subtraction term $\iota_{\mathbf{q}}$ does not depend on $q^{0}$ and therefore drops out in the momentum-projected correlator $\tilde{c}_{SS}(t>0,\mathbf{p})$,
\begin{eqnarray}
\tilde{c}_{SS}^{fv}(t,\mathbf{p})|_{\phi\phi-\mathrm{ex.}} &=& \frac{b^2}{\pi}\int_{4M^2}^{\infty}ds'\left(\sum_{\mathbf{n}\in\mathds{Z}^3}\frac{\pi\left(E_{+}^{(\mathbf{n})}+E_{-}^{(\mathbf{n})}\right)}{4L^3E_{+}^{(\mathbf{n})}E_{-}^{(\mathbf{n})}}\delta\left(s'-s_{\mathbf{n}}^{(0)}(\mathbf{p})\right)\right)\frac{e^{-i\sqrt{s'+|\mathbf{p}|^2}t}}{2\sqrt{s'+|\mathbf{p}|^2}} \nonumber \\
 &=& \frac{b^2}{2L^3}\sum_{\mathbf{n}\in\mathds{Z}^3}\frac{e^{-i\left(E_{+}^{(\mathbf{n})}+E_{-}^{(\mathbf{n})}\right)t}}{(2E_{+}^{(\mathbf{n})})(2E_{-}^{(\mathbf{n})})}\,,\label{eq:fifi2fv}
\end{eqnarray}
with $\mathbf{p}=\frac{2\pi}{L}\mathbf{N}$, $\mathbf{N}\in\mathds{Z}^{3}$. It should be clear that Eqs.~(\ref{eq:fifi2}) and (\ref{eq:fifi2fv}) just correspond to an insertion of a complete set of {\em free\,} $\phi\phi$ states. Schematically, $\tilde{c}_{SS}(t,\mathbf{p})|_{\phi\phi-\mathrm{ex.}}$ is expanded as
\begin{equation}
\quad\,\sim\, \int d^{3}x\,e^{-i\mathbf{p}\cdot\mathbf{x}}\frac{1}{2}\int\frac{d^{3}q_{1}}{(2\pi)^3}\frac{1}{2E_{q_{1}}}\int\frac{d^{3}q_{2}}{(2\pi)^3}\frac{1}{2E_{q_{2}}}\langle 0|\mathcal{S}(x)|\phi(\mathbf{q}_{1})\phi(\mathbf{q}_{2})\rangle\langle\phi(\mathbf{q}_{1})\phi(\mathbf{q}_{2})|\mathcal{S}(0)|0\rangle\,.
\end{equation}
\newpage
\underline{{\em Remark on Eq.~(\ref{eq:ifififv1}):}}\\
Note that the limit $\epsilon\rightarrow 0+$ is implicit in Eq.~(\ref{eq:ifififv1}). It is obvious that the sum of (real) pole terms cannot be a meaningful approximation to the loop integral for $q^2>4M^2$ in this limit, which is complex for such $q^2$, and has no poles. This is because the integrand has a singularity in the integration range (see e.g. Eq.~(\ref{eq:fifi1})). In other words, the limits $(2\pi/L)\rightarrow 0$ and $\epsilon\rightarrow 0$ do not commute. To isolate the problematic part, we subtract a certain sum of pole terms from those of Eq.~(\ref{eq:ifififv1}), with coefficients chosen such that, when $q^2$ comes close to a certain pole position with index $\mathbf{n}$, and $\epsilon\rightarrow 0$, the corresponding pole cancels out. To this end, we note the following integral representation of the imaginary part of $I_{\phi\phi}(s)$, for $s>4M^2$:
\begin{equation}
\frac{1}{\sqrt{s+i\epsilon}}\int\frac{d^{3}\mathbf{l}}{(2\pi)^3}\left(\frac{1}{s+i\epsilon-4(|\mathbf{l}|^2+M^2)} + \frac{|\mathbf{l}|^2+3M^2}{4(|\mathbf{l}|^2+M^2)^2}\right) = -\frac{i\sigma(s+i\epsilon)}{32\pi}\,,\label{eq:isigmaInt}
\end{equation}
which suggests to subtract the corresponding terms in the integrand from $\bar{I}_{\phi\phi}^{fv}$ (let us set $\mathbf{N}=\mathbf{0}$ for this demonstration):
\begin{eqnarray}
p_{\epsilon}(s,L) &:=& \left(\frac{2\pi}{L}\right)^3\sum_{\mathbf{n}\in\mathds{Z}^3}\left(\frac{1}{16\pi^3E_{0}^{(\mathbf{n})}(s+i\epsilon-4(E_{0}^{(\mathbf{n})})^2)} + \frac{1}{64\pi^3(E_{0}^{(\mathbf{n})})^3} \right) \nonumber \\ &-& \left(\frac{2\pi}{L}\right)^3\sum_{\mathbf{n}\in\mathds{Z}^3}\left(\frac{1}{8\pi^3\sqrt{s+i\epsilon}\,(s+i\epsilon-4(E_{0}^{(\mathbf{n})})^2)} + \frac{1}{8\pi^3}\frac{(E_{0}^{(\mathbf{n})})^2+2M^2}{4\sqrt{s+i\epsilon}\,(E_{0}^{(\mathbf{n})})^4}\right) \nonumber \\
 &=& \left(\frac{2\pi}{L}\right)^3\sum_{\mathbf{n}\in\mathds{Z}^3}\frac{(s+i\epsilon-8M^2)E_{0}^{(\mathbf{n})}-4M^2\sqrt{s+i\epsilon}}{64\pi^3(E_{0}^{(\mathbf{n})})^4\sqrt{s+i\epsilon}\,(\sqrt{s+i\epsilon}+2E_{0}^{(\mathbf{n})})}\,. %
\end{eqnarray}
This expression has no pole in $s$ for $s>4M^2$, so we can take the limit $\epsilon\rightarrow 0,\,L\rightarrow\infty$ to obtain the integral
\begin{eqnarray}
p_{\epsilon}(s,L) &\rightarrow& p_{0}(s,\infty) = \frac{1}{\sqrt{s}}\int\frac{d^{3}\mathbf{l}}{(2\pi)^3}\frac{(s-8M^2)\sqrt{|\mathbf{l}|^2+M^2}-4M^2\sqrt{s}}{8(\sqrt{|\mathbf{l}|^2+M^2})^4(\sqrt{s}+2\sqrt{|\mathbf{l}|^2+M^2})} \nonumber \\
 &=& \frac{1}{16\pi^2}\left(\sigma(s)\artanh(\sigma(s))-1\right) = \mathrm{Re}\,\bar{I}_{\phi\phi}(s)-\frac{1}{16\pi^2}\,.
\end{eqnarray}
So we have learnt that we can approximate (neglecting in particular exponentially suppressed terms $\sim K_{0}\left(ML|\mathbf{k}|\right)$ in Eq.~(\ref{eq:ifififv1})) for $s>4M^2$, $ML\gg 1\,$:
\begin{equation}
\bar{I}^{fv}_{\phi\phi}(s,L) \approx  \mathrm{Re}\,\bar{I}_{\phi\phi}(s) \,+\, \frac{1}{\sqrt{s}L^3}\sum_{\mathbf{n}\in\mathds{Z}^3}\left(\frac{1}{s-4(E_{0}^{(\mathbf{n})})^2} + \frac{(E_{0}^{(\mathbf{n})})^2+2M^2}{4(E_{0}^{(\mathbf{n})})^4}\right) \,.\label{eq:ifififv1approx}
\end{equation}
It is (only) in the sense of Eqs.~(\ref{eq:isigmaInt}) and (\ref{eq:ifififv1approx}) that $\bar{I}^{fv}_{\phi\phi}$ is a finite-volume ``approximation'' to $\bar{I}_{\phi\phi}$ for $s>4M^2$ (for $s<4M^2$, we have $\bar{I}^{fv}_{\phi\phi}\approx\bar{I}_{\phi\phi}$).
\newpage
\section{Unitary model for the scattering amplitude}
\label{sec:T}
In this section, we describe our model for the $\sigma$ self-energy, the $\sigma\phi\phi$ vertex function and the $\phi\phi$ scattering amplitude. We split the complete scattering amplitude as
\begin{displaymath}
T=T_{p}+T_{np}\,,
\end{displaymath}
where $T_{p}$ collects all terms containing the resonance pole, i.e. all s-channel resonance exchange graphs, while $T_{np}$ contains the non-resonant ``background''. Our ansatz for $T_{np}$ is
\begin{equation}
T_{np}(s)=\frac{1}{\alpha^{-1}+\bar{I}_{\phi\phi}(s)}\,,
\end{equation}
with a real parameter $\alpha$ and the loop function (see Eq.~(\ref{eq:fifi1}))
\begin{equation}\label{eq:fifi1explicit}
\bar{I}_{\phi\phi}(s) = -\frac{\sigma(s)}{16\pi^2}\artanh\left(-\frac{1}{\sigma(s)}\right)\,,\quad \sigma(s)=\sqrt{1-\frac{4M^2}{s}}\,.
\end{equation}
This amplitude has no resonance poles by construction. For $\alpha<0$, however, there are ``artefact'' poles on the negative $s$-axis of the physical sheet, which have no physical interpretation. We can use the background amplitude to construct the $\sigma\phi\phi$ vertex function,
\begin{equation}
\Gamma(s)=g-g\bar{I}_{\phi\phi}(s)T_{np}=\frac{g}{1+\alpha \bar{I}_{\phi\phi}(s)}\,,
\end{equation}
with a real resonance coupling parameter $g$, and the self-energy of the $\sigma$ resonance
\begin{equation}
\Sigma(s)=g\bar{I}_{\phi\phi}(s)\Gamma(s)\,.
\end{equation}
This self-energy obeys the unitarity relation (for real $s>4M^2$)
\begin{equation}\label{eq:sigmauni}
\mathrm{Im}\,\Sigma(s)=\Gamma^{\ast}(s)(\mathrm{Im}\,\bar{I}_{\phi\phi}(s))\Gamma(s) = -\Gamma^{\ast}(s)\left(\frac{\sigma(s)}{32\pi}\right)\Gamma(s)\,.
\end{equation}
The pole-part of the scattering amplitude is then constructed as
\begin{equation}\label{eq:Tpolpart}
T_{p}(s)=-\Gamma(s)\frac{1}{s-m_{0}^{2}-\Sigma(s)}\Gamma(s)\,,
\end{equation}
and the full scattering amplitude can simply be written as
\begin{equation}\label{eq:fullT}
T(s) = \frac{1}{\left(\alpha+\frac{g^2}{m_{0}^{2}-s}\right)^{-1}+\bar{I}_{\phi\phi}(s)} = \frac{1}{\left(\alpha+\frac{g^2}{m_{0}^{2}-4M^2}\right)^{-1}+\frac{s-4M^2}{s_{2}-4M^2}\frac{g^2/\alpha^2}{s-s_{2}}+\bar{I}_{\phi\phi}(s)}\,,
\end{equation}
with $s_{2}:=m_{0}^{2}+(g^2/\alpha)$. The mass parameter $m_{0}$ can be interpreted as the mass of the ``undressed'' resonance. 
The model described above is very simple - it has only one more parameter than a Breit-Wigner parameterization. Moreover, it is a pure s-wave amplitude, with the s-wave phase shift $\delta_{0}(s)$ given by
\begin{equation}\label{eq:fullTphase}
t_{0}(s)=\frac{1}{32\pi}T(s)=\frac{e^{2i\delta_{0}(s)}-1}{2i\sigma(s)}\,,\quad s>4M^2\,.
\end{equation}
The model amplitude fulfills elastic unitarity exactly, $\mathrm{Im}\,t_{0}=t_{0}^{\ast}\sigma(s)t_{0}$ for $s>4M^2$, but all higher-lying (multi-) particle intermediate states are neglected, so it should be viewed as an effective description valid only below inelastic thresholds. It also does not possess a left-hand cut. \\
\begin{itemize}
 \item For $\alpha>0,\,m_{0}>2M$, the amplitude has a resonance pole on its second Riemann sheet, and is analytic on the cut physical sheet. It is then a solution to the ``one-channel Roy equation'' \cite{Gasser:1999hz} with one CDD pole (\cite{Castillejo:1955ed}, see also Sec.~3 of \cite{Brander:1975ed}).
 \item For $\alpha>0,\,m_{0}<2M$, there is in addition a bound state at some energy $0<\sqrt{s}<2M$.
 \item For $\alpha<0$, there are unphysical poles on the first Riemann sheet. This is unpleasant, but since these poles could occur at large $|s|$, such a result could not rule out the present model as a valid description at low energies. Should such a pole, however, occur in the low-energy region (say, for $|s|<16M^2$), the solution should be rejected.
\end{itemize}
Recall that two additional parameters $b$ and $\tilde{b}$ are introduced, which give the direct coupling of the scalar operator to two free $\phi$ fields and to an undressed $\sigma$, respectively. \\
\section{Unitary model for the correlator}
\label{sec:M}
From the simple model for the $\phi\phi$ scattering amplitude described in the previous section, we can construct a model for the Fourier transform $\mathcal{M}$ of the correlator. Let
\begin{eqnarray}
\beta(s) &=& b\left(1-\bar{I}_{\phi\phi}(s)T_{np}(s)\right)\,,\\
\tilde{\beta}(s) &=& \tilde{b} - b\bar{I}_{\phi\phi}(s)\Gamma(s)\,.
\end{eqnarray}
Then the contributions with ($p$) and without ($np$) resonance pole terms are
\begin{equation}
\mathcal{M}_{p}(s) = \frac{\tilde{\beta}^{2}(s)}{s-m_{0}^{2}-\Sigma(s)}\,,\qquad \mathcal{M}_{np}(s) = b\bar{I}_{\phi\phi}(s)\beta(s)\,, \qquad \mathcal{M}(s) = \mathcal{M}_{p}(s) + \mathcal{M}_{np}(s)\,.
\end{equation}
It is then straightforward to show that, for real $s=q^2$,
\begin{eqnarray}
\mathcal{M}(s) &=& \frac{b^2\bar{I}_{\phi\phi}(s) + \frac{\tilde{b}^2(1+\alpha\bar{I}_{\phi\phi}(s))  -2\tilde{b}\bar{I}_{\phi\phi}(s)gb}{s-m_{0}^{2}}}{1+\bar{I}_{\phi\phi}(s)\left(\alpha+\frac{g^2}{m_{0}^{2}-s}\right)}\,,\quad \mathcal{M}(m_{0}^{2})  = \frac{\tilde{b}}{g^2\bar{I}_{\phi\phi}(m_{0}^{2})}\left(\tilde{b}+\bar{I}_{\phi\phi}(m_{0}^{2})\left(\alpha\tilde{b}-2gb\right)\right)\,,\nonumber\\
\mathrm{Im}\,\mathcal{M}(s) &=& B^{\ast}(s)\left(\mathrm{Im}\bar{I}_{\phi\phi}(s)\right)B(s) \,,\qquad
B(s) = \frac{b+\frac{\tilde{b}g}{m_{0}^{2}-s}}{1+\bar{I}_{\phi\phi}(s)\left(\alpha+\frac{g^2}{m_{0}^{2}-s}\right)}\,.\label{eq:fullM}
\end{eqnarray}
Note that $B(s)$ and $\mathcal{M}(s)$ have the same pole structure as $T(s)$ of Eq.~(\ref{eq:fullT}), so that the constraints of Eq.~(\ref{eq:dispM}) are fulfilled for our model amplitude (given that $\alpha>0,\,m_{0}>2M$).\\
Our Eqs.~(\ref{eq:dispM}) and (\ref{eq:cfull}) correspond to a summation over $\phi\phi$ intermediate states: There are no ``$\sigma$ states'' in the case where the $\sigma$ is a resonance. The resonance effects are all contained in the pole structure of the complex function $B(s)$, which is the same as that of $T(s)$ (see Eq.~(\ref{eq:fullT})),
\begin{equation}\label{eq:Buni}
B(s)=B_{0}(s)-B_{0}(s)\bar{I}_{\phi\phi}(s)T(s)\,,\quad B_{0}(s):= b+\frac{\tilde{b}g}{m_{0}^{2}-s}\,,
\end{equation}
so that, from (\ref{eq:fifi1explicit}), (\ref{eq:fullTphase}) and (\ref{eq:Buni}), one can show $B(s)=|B(s)|e^{i\delta_{0}(s)}$ (``Watson theorem''). \\
In the case where the $\sigma$ is a bound state, the amplitude $\mathcal{M}_{p}(s)$ has a pole at $s=s_{\sigma}\overset{!}{=}m_{\sigma}^{2}$ on the {\em first\,} sheet, which therefore directly appears in the spectrum, just like the ``undressed'' resonance would (compare Eq.~(\ref{eq:baresigmaex})),
\begin{eqnarray}
\mathcal{M}_{p}(s) &=& \frac{Z_{\sigma}\tilde{\beta}^{2}(s_{\sigma})}{s-s_{\sigma}}+\ldots \,,\qquad Z_{\sigma}:=\frac{1}{1-\Sigma'(s_{\sigma})}\,,\label{eq:polZsigma}\\
 \tilde{c}_{SS}(t,\mathbf{p})|_{\sigma_{b}-\mathrm{ex.}} &=& \frac{Z_{\sigma}\tilde{\beta}^{2}(m_{\sigma}^{2}) e^{-it\sqrt{|\mathbf{p}|^2+m_{\sigma}^{2}}}}{2\sqrt{|\mathbf{p}|^2+m_{\sigma}^{2}}}\,\qquad (\sigma_{b}:\,\textrm{bound state})\,.
\end{eqnarray}
In the resonant case, the pole is located on the {\em second\,} Riemann sheet, at $s_{\sigma}=\left(m_{\sigma}-(i/2)\Gamma_{\sigma}\right)^2$, and therefore does not appear in the dispersive representation of the amplitude $\mathcal{M}$ (compare Eqs.~(\ref{eq:dispM}), (\ref{eq:cfull})). What can be extracted from the time-dependence of the correlator is only the function $|B(s)|$ (written as $|B(E')|$ in Eq.~(\ref{eq:cfull}), where $\mathbf{p}$ is fixed). With
\begin{equation}\label{eq:polcond}
s_{\sigma}-m_{0}^{2}-\Sigma(s_{\sigma})\overset{!}{=}0\,\Rightarrow\,\bar{I}_{\phi\phi}(s_{\sigma})=\frac{s_{\sigma}-m_{0}^{2}}{g^2-\alpha(s_{\sigma}-m_{0}^{2})}\,
\end{equation}
(note that $\bar{I}_{\phi\phi}$ is to be evaluated on the second sheet here), it is possible to show that 
\begin{equation}\label{eq:resBoverressqrtT}
\sqrt{Z_{\sigma}}\tilde{\beta}(s_{\sigma}) = -\mathrm{lim}_{s\rightarrow s_{\sigma}}\frac{(s-s_{\sigma})B(s)}{\sqrt{Z_{\sigma}}\Gamma(s)} = -\frac{\mathrm{Res}_{s_{\sigma}}B}{\sqrt{-\mathrm{Res}_{s_{\sigma}}T}}\,,
\end{equation}
which gives us (within our present model) the generalization of the matrix element $\langle 0|\mathcal{S}(0)|\sigma\rangle$ for the case where the $\sigma$ is a resonance, in terms of the residues of the form factor $B$ and the scattering amplitude $T$ at the resonance pole.
\newpage
\subsection*{Finite-volume analysis}
In a finite volume $L^3$, the loop function $\bar{I}_{\phi\phi}$ in (\ref{eq:fullM}) has to be replaced by its finite-volume counterpart $\bar{I}_{\phi\phi}^{fv}(q,L)$ (see Eq.~(\ref{eq:ifififv1})), and we deduce that $\mathcal{M}(q)$ is then given as a series of {\em pole\,} terms,
\begin{equation}\label{eq:Mpolform}
\mathcal{M}(q,L) = \mathrm{const.}\, + 4\left(\frac{2\pi}{L}\right)^2\sum_{j=0}^{\infty}\frac{r_{j}}{q^2-s_{j}}\,,\quad s_{j+1}>s_{j}\,,
\end{equation}
where the pole positions are shifted from the values $s_{\mathbf{n}}^{(0)}$ (see Eq.~(\ref{eq:ifififv2})) by the interaction, $s_{j(\mathbf{n})}=s_{\mathbf{n}}^{(0)}+\mathcal{O}(\alpha,g^2)\,$. In principle, the form Eq.~(\ref{eq:Mpolform}) is just given by the discretization of the dispersive integral in Eq.~(\ref{eq:dispM}), which should be a valid approximation to this integral at distances $\gg(2\pi/L)^2$ from the positive real $s$-axis.\\
In the present model, it is easy to see that there are only {\em simple\,} poles: Suppose that there are two neighboring pole positions $s_{1,2}$ of $\mathcal{M}(q,L)$, $s_{2}=s_{1}+\delta s>s_{1}$, and apply the pole conditions of Eq.~(\ref{eq:polcond}),
\begin{eqnarray}
\bar{I}_{\phi\phi}^{fv}(s_{2})-\bar{I}_{\phi\phi}^{fv}(s_{1}) &\overset{!}{=}& \frac{(s_{2}-s_{1})g^2}{(g^2-\alpha(s_{2}-m_{0}^{2}))(g^2-\alpha(s_{1}-m_{0}^{2}))}\nonumber \\
\Rightarrow\quad \frac{d}{ds}\bar{I}_{\phi\phi}^{fv}(s_{1})\delta s + \mathcal{O}((\delta s)^2) &\overset{!}{=}& \delta s \frac{g^2}{(g^2-\alpha(s_{1}-m_{0}^{2}))^2}+\mathcal{O}((\delta s)^2)\,,
\end{eqnarray}
but the leading term on the right-hand side is non-negative, while the one on the left-hand side is non-positive (considering Eq.~(\ref{eq:ifififv1}) with $q^2\rightarrow s$), and thus $\delta s$ cannot become arbitrarily small as long as $L<\infty$, $g^2>0$ (``avoided level crossing''). \\
It is well-known \cite{Luscher:1990ux} that the spectrum given by the $s_{j}$ determines a finite-volume approximation to the (infinite-volume) phase-shift $\delta_{0}(s)$ at $s=s_{j}$. This is easily seen for toy models as discussed here, with simple real potentials like $V(s)=\alpha+(g^2/(m_{0}^{2}-s))$, where the phase-shift is explicitly given by (see Eqs.~(\ref{eq:fullT}), (\ref{eq:fullTphase}))
\begin{equation}\label{eq:exp2idelta}
e^{2i\delta_{0}(s)}=\frac{32\pi\left(V^{-1}(s)+\mathrm{Re}\,\bar{I}_{\phi\phi}(s)\right)+i\sigma(s)}{32\pi\left(V^{-1}(s)+\mathrm{Re}\,\bar{I}_{\phi\phi}(s)\right)-i\sigma(s)} \equiv \frac{\cot\delta_{0}(s)+i}{\cot\delta_{0}(s)-i}\,.
\end{equation}
Neglecting all finite-volume effects which are exponentially damped with $ML$, and writing the equation determining the pole positions in the finite volume (see Eq.~(\ref{eq:polcond})) as
\begin{equation}\label{eq:polcond_rewritten}
V^{-1}(s_{j})+\bar{I}_{\phi\phi}^{fv}(s_{j})\overset{!}{=}0\,,
\end{equation}
we can approximate the expression for the phase shift using Eq.~(\ref{eq:ifififv1approx}),
\begin{equation}\label{eq:useifififv1approx}
V^{-1}(s_{j})+\mathrm{Re}\,\bar{I}_{\phi\phi}(s_{j}) \approx -\frac{1}{\sqrt{s_{j}}L^3}\sum_{\mathbf{n}\in\mathds{Z}^3}\left(\frac{1}{s_{j}-4(E_{0}^{(\mathbf{n})})^2} + \frac{(E_{0}^{(\mathbf{n})})^2+2M^2}{4(E_{0}^{(\mathbf{n})})^4}\right)\,,\quad j=0,1,2,\ldots
\end{equation}
\newpage
In the continuum limit, the pole positions $s_{i}$ move closer and closer together, and the function $\mathcal{M}(q)$ acquires a branch cut, and a Riemann sheet structure. The function is then essentially determined by a spectral function, integrated along the unitarity branch cut, as in Eq.~(\ref{eq:dispM}). The question is whether this spectral function can somehow be (approximately) reconstructed from the set of numbers $\left\{r_{i},s_{i}\right\}$, in a way similar to the reconstruction of the scattering phase shift with the help of the (squared) energy levels $s_{i}$ and Eqs.~(\ref{eq:exp2idelta})-(\ref{eq:useifififv1approx}).\\
From (\ref{eq:Mpolform}) and (\ref{eq:cp}), (\ref{eq:cM}), the momentum-projected correlator in a finite volume will be of the form
\begin{equation}\label{eq:cMpfv}
\tilde{c}_{SS}^{fv}(t,\mathbf{p}) = 4\left(\frac{2\pi}{L}\right)^2\sum_{j}\frac{r_{j}\,e^{-i\sqrt{s_{j}+|\mathbf{p}|^2}t}}{2\sqrt{s_{j}+|\mathbf{p}|^2}}\,.
\end{equation}
Assume that the poles are distributed as $s_{j+1}-s_{j}=4\left(\frac{2\pi}{L}\right)^2/\rho(s_{j})$, with a positive real function $\rho(s)$. 
We can apply a rescaling function to obtain a set of positions $\tilde{s}_{j}$ with an equidistant distribution, $\tilde{s}(s_{j})=s_{0}+4j\left(\frac{2\pi}{L}\right)^2$ (the prefactors here and in Eqs.~(\ref{eq:Mpolform}), (\ref{eq:cMpfv}) have been chosen such that $\rho(s)=1$ for the {\em free\,} case, with $\mathbf{N}=\mathbf{0}$). In the infinite-volume limit, the sum in Eq.~(\ref{eq:cMpfv}) tends to the integral
\begin{equation}
\tilde{c}_{SS}^{fv}(t,\mathbf{p})\,\rightarrow\,\int_{4M^2}^{\infty}d\tilde{s}\,\frac{r(s')\,e^{-i\sqrt{s'+|\mathbf{p}|^2}\,t}}{2\sqrt{s'+|\mathbf{p}|^2}} = \int_{4M^2}^{\infty}ds'\,\rho(s')\frac{r(s')\,e^{-i\sqrt{s'+|\mathbf{p}|^2}\,t}}{2\sqrt{s'+|\mathbf{p}|^2}}\,,
\end{equation}
where $r(s')$ is a smooth function that interpolates between the values $r_{j}$. The last equation has to be compared with Eqs.~(\ref{eq:fifi2}) and (\ref{eq:cfull}). With the help of those equations, together with (\ref{eq:fullM}) (with $\bar{I}_{\phi\phi}$ replaced by $\bar{I}_{\phi\phi}^{fv}(q,L)$ for the finite volume), and (\ref{eq:polcond_rewritten}), we find (recall that $V(s):=\alpha+(g^2/(m_{0}^{2}-s))\,$):
\begin{eqnarray}
\rho(s_{j}) &\overset{!}{\approx}& -\left(\frac{\sigma(s_{j})}{32\pi^2}\right)\frac{4\left(\frac{2\pi}{L}\right)^2V(s_{j})\frac{d}{ds}\left[V(s)\bar{I}_{\phi\phi}^{fv}(s)\right]_{s_{j}}}{\left|1+V(s_{j})\bar{I}_{\phi\phi}(s_{j})\right|^2}\nonumber \\
 &=& -\left(\frac{\sigma(s_{j})}{32\pi^2}\right)\frac{4\left(\frac{2\pi}{L}\right)^2\frac{d}{ds}\left[(V(s))^{-1}+\bar{I}_{\phi\phi}^{fv}(s)\right]_{s_{j}}}{\left|(V(s_{j}))^{-1}+\bar{I}_{\phi\phi}(s_{j})\right|^2}\,.\label{eq:rhobycomparison}
\end{eqnarray}
Note that the last expression is {\em positive\,} because both $\frac{d}{ds}V^{-1}$ and $\frac{d}{ds}\bar{I}_{\phi\phi}^{fv}$ are negative. From Eq.~(\ref{eq:ifififv1approx}), we can express the derivative in the numerator as
\begin{eqnarray}\label{eq:dcot}
\frac{d}{ds}\left[(V(s))^{-1}+\bar{I}_{\phi\phi}^{fv}(s)\right]_{s_{j}} &\approx& \frac{d}{ds}\left[(V(s))^{-1}+\mathrm{Re}\,\bar{I}_{\phi\phi}(s)\right]_{s_{j}}+\frac{1}{2s_{j}}\left((V(s_{j}))^{-1}+\mathrm{Re}\,\bar{I}_{\phi\phi}(s_{j})\right) \nonumber \\ &-& \frac{1}{\sqrt{s_{j}}L^3}\sum_{\mathbf{n}\in\mathds{Z}^3}\frac{1}{\left(s_{j}-4(E_{0}^{(\mathbf{n})})^2\right)^2}\,.
\end{eqnarray}
For $\mathbf{p}\not=0$, the corresponding energy levels should be used. Inserting (\ref{eq:dcot}) in (\ref{eq:rhobycomparison}), and using Eq.~(\ref{eq:exp2idelta}) and its derivative w.r.t. $s$, we find that the spectral function of $\mathcal{M}$ (see Eqs.~(\ref{eq:dispM}), (\ref{eq:fullM})) can be reconstructed from the measured matrix elements in the finite volume in the following sense:
\begin{equation}\label{eq:spectralf}
\left(\frac{\sigma(s_{j})}{32\pi}\right)|B(s_{j})|^2 \approx \frac{\pi\rho(s_{j})}{4\left(\frac{2\pi}{L}\right)^2}\left[4\left(\frac{2\pi}{L}\right)^2\,r_{j}\right]\,,
\end{equation}
with
\begin{equation}\label{eq:rhosol}
\frac{\pi\rho(s_{j})}{4\left(\frac{2\pi}{L}\right)^2} \approx \frac{d\delta_{0}(s)}{ds}\biggr|_{s_{j}} - \frac{\sin 2\delta_{0}(s_{j})}{4(s_{j}-4M^2)} + \frac{32\pi}{\sqrt{s_{j}-4M^2}L^3}\sum_{\mathbf{n}\in\mathds{Z}^3}\frac{\sin^{2}\delta_{0}(s_{j})}{\left(s_{j}-4(E_{0}^{(\mathbf{n})})^2\right)^2} \,.
\end{equation}
The requirement that 
\begin{equation}\label{eq:wignerbound}
\frac{d\delta_{0}(s)}{ds} - \frac{\sin 2\delta_{0}(s)}{4(s-4M^2)} \overset{!}{>} 0 
\end{equation}
is known as Wigner's causality bound for an s-wave zero-range potential, see Eq.~(5a) in \cite{Wigner:1955zz}.
 - The prefactors of $|B(s_{j})|^2$ and $4\left(\frac{2\pi}{L}\right)^2r_{j}$ in Eq.~(\ref{eq:spectralf}) have the simple interpretation of energy-level densities. \\
To conclude, one can extract resonance properties from numerical finite-volume data as follows:
\begin{itemize}
 \item Determine the (squared) energy levels $s_{i}$ and the coefficients $4\left(\frac{2\pi}{L}\right)^2r_{j}$ which determine the momentum-projected correlator (Eq.~(\ref{eq:cMpfv}) evaluated in euclidean time $t=-i\tau$), 
 \item reconstruct the scattering phase shift in the infinite volume via Eq.~(\ref{eq:exp2idelta}) (with (\ref{eq:useifififv1approx}))\,,
 \item use a parameterization of the scattering amplitude, which satisfies the correct analyticity and unitarity conditions, to make a fit to the phase-shift ``data'', and determine the mass and width of the resonance (given by the resonance pole position on the second Riemann sheet) by analytic continuation of $T(s)$ to the complex $s$-surface.
 \item Use the energy levels and the scattering phase shift $\delta_{0}(s)$ to compute the density function $\rho(s)$ (Eq.~(\ref{eq:rhosol})).
 \item Obtain $|B(s_{j})|^2$ from the measured coefficients $4\left(\frac{2\pi}{L}\right)^2r_{j}$ via Eq.~(\ref{eq:spectralf}).
 \item Obtain the time-like form factor $B(s)=|B(s)|e^{i\delta_{0}(s)}$ for $s\geq 4M^2$.
 \item Employ a parameterization for $B(s)$ to analytically continue to the complex plane, as done for the scattering amplitude $T(s)$, and find the resonance decay matrix element from the corresponding residues at the resonance pole (Eq.~(\ref{eq:resBoverressqrtT})).
\end{itemize}
We will demonstrate this procedure in the next section.
\newpage
\section{Numerical demonstration}
\label{sec:Num}
Our field-theoretical model for the scattering amplitude and the correlator is specified in Eqs.~(\ref{eq:fullT}) and (\ref{eq:fullM}). We shall employ units such that the mass of the $\phi$ particles is $M=1$. In these units, the ``true values'' of the five model parameters will be taken as
\begin{equation}\label{eq:truepars}
\alpha=25\,,\quad g=15\,, \quad m_{0}=3\,,\quad b=1\,,\quad \tilde{b}=\frac{2}{3}\,.
\end{equation}
For these values, we find that the resonance pole of $T(s)$ on the second sheet is located at
\begin{displaymath}
s_{\sigma}=\left(m_{\sigma}-\frac{i}{2}\Gamma_{\sigma}\right)^2\,,\quad m_{\sigma}=3.211\,,\quad \Gamma_{\sigma}=0.457\,,
\end{displaymath}
with residuum $\mathrm{Res}_{s_{\sigma}}T=-176.26-71.77\,i$. \\
Let us assume that the correlator $\tilde{c}_{SS}(t,\mathbf{0})$ is measured in a numerical simulation in a finite volume with $ML=10$, with data given in the table below:
\begin{center}
\vspace{-0.5cm}
\begin{table}[h]
\label{tab:data}
\begin{tabular}{|c||c|c|c|c|c|c|c|c|c|c|c|c|c|c|c|c|}
\hline
$j$ & 0 & 1 & 2 & 3 & 4 & 5 & 6 & 7  \\
\hline
$s_{j}$ & 3.971 & 5.398 & 6.818 & 8.447 & 9.718 & 10.587 & 12.244 & 13.669 \\
\hline
$4\left(\frac{2\pi}{L}\right)^2r_{j}$ & 0.003 & 0.023 & 0.042 & 0.048 & 0.122 & 0.053 & 0.030 & 0.012 \\
\hline %
\end{tabular}%
\end{table}%
\end{center}%
\vspace{-1cm}
We do not attempt any error analysis here - the present section should just serve to demonstrate the plausibility and applicability of our foregoing work. From the values $s_{j}$, we can evaluate the phase shift $\delta_{0}(s)$ at the points $s=s_{j}$ in the low-energy region $4M^2<s<16M^2$. 
\begin{center}
\begin{figure}[h]
\includegraphics[width=8.5cm]{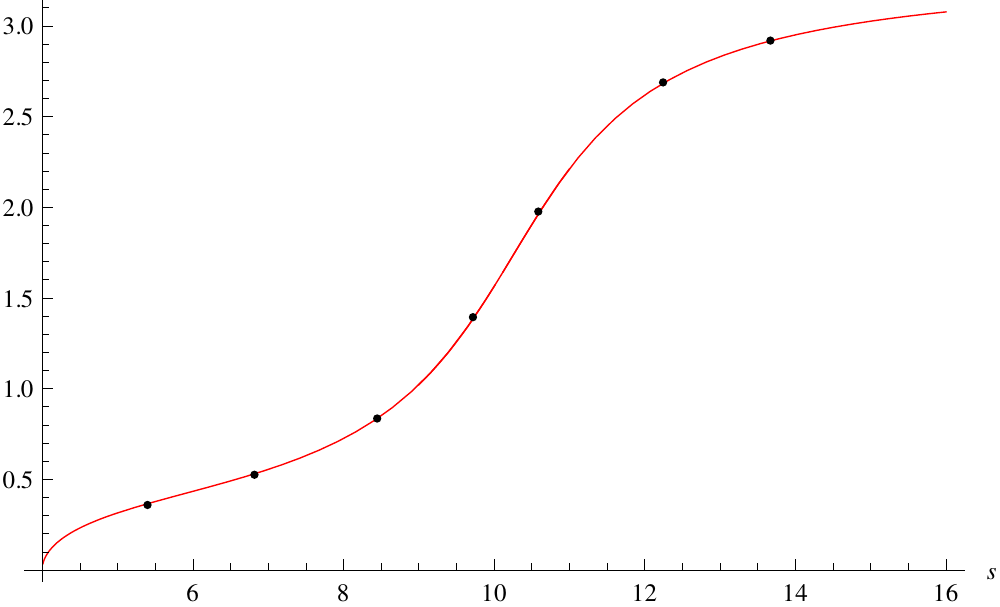}
\end{figure}
\end{center}%
\vspace{-1cm}
The plot shows that Eqs.~(\ref{eq:exp2idelta}), (\ref{eq:useifififv1approx}) work nicely for $ML=10$. We also see that there is a clear signature of the resonance in the phase shift. The red curve shows the exact phase shift in the infinite volume, while the black dots indicate the values computed with our finite-volume formulae.\\
We can now make a fit of some parameterization for the scattering amplitude to the phase-shift ``data'' points. Using our model amplitude of Eq.~(\ref{eq:fullT}), for example, a fit to the data points in $\left[4M^2,16M^2\right]$ returns
\begin{displaymath}
\alpha_{fit}=24.28\,,\quad g_{fit}=14.87\,, \quad m_{0,fit}=3.00\,,
\end{displaymath}
which compared to the ``true'' values given above is an extremely good result. Analytically continuing to the second Riemann sheet, we find a pole located at
\begin{displaymath}
 m_{\sigma,fit}=3.207\,,\quad \Gamma_{\sigma,fit}=0.453\,,\quad \mathrm{Res}_{s_{\sigma}}T_{fit}=-175.44-69.16\,i\,.
\end{displaymath}
Of course, one does not know the true form of the scattering amplitude in practice. Let us try a Breit-Wigner-like parameterization (compare (\ref{eq:sigmauni}), (\ref{eq:Tpolpart}), (\ref{eq:fullTphase})),
\begin{equation}\label{eq:BW}
T_{BW}(s) := -\frac{\gamma_{BW}^2}{s-m_{BW}^2+i\gamma_{BW}^2\left(\frac{\sigma(s)}{32\pi}\right)}\,.
\end{equation}
This would give $m_{BW}=2.97$, $\gamma_{BW}=13.88$, and a pole with $m_{\sigma,BW}=2.960$, $\Gamma_{\sigma,BW}=0.482$.\\
Using the parameterization $T(s)$ of Eq.~(\ref{eq:fullT}) and the values $\alpha_{fit},\,g_{fit},\,m_{0,fit}$, we can now evaluate the numbers $\rho(s_{j})$ with the help of Eq.~(\ref{eq:rhosol})\,:
\begin{center}
\vspace{-0.5cm}
\begin{table}[h]
\label{tab:rhodata}
\begin{tabular}{|c||c|c|c|c|c|c|c|c|c|c|c|c|c|c|c|c|}
\hline
$j$  & 1 & 2 & 3 & 4 & 5 & 6 & 7  \\
\hline
$\rho(s_{j})$ & 1.011 & 0.884 & 1.557 & 1.055 & 2.106 & 0.878 & 0.579 \\
\hline %
\end{tabular}%
\end{table}%
\end{center}%
\vspace{-1cm}
The lowest level with $j=0$ is excluded here because $\delta_{0}(s)$ is only defined for $s\geq 4M^2$.
Now we are in a position to compare the outcome for the right-hand side of Eq.~(\ref{eq:spectralf}) (numerical results) with the ``true'' values on the left-hand side, given by our model for $|B(s_{j})|^2$ and the ``true'' parameters of Eq.~(\ref{eq:truepars}):
\begin{center}
\vspace{-0.5cm}
\begin{table}[h]
\label{tab:specdata}
\begin{tabular}{|c||c|c|c|c|c|c|c|c|c|c|c|c|c|c|c|c|}
\hline
$j$  & 1 & 2 & 3 & 4 & 5 & 6 & 7  \\
\hline
LHS of Eq.~(\ref{eq:spectralf}) (``true'') & \,0.047\, & \,0.076\, & \,0.149\, & \,0.254\, & \,0.225\, & \,0.053\, & \,0.014\, \\
\hline
RHS of Eq.~(\ref{eq:spectralf}) (``num.'') & 0.045 & 0.074 & 0.148 & 0.255 & 0.222 & 0.052 & 0.014 \\
\hline %
RHS of Eq.~(\ref{eq:spectralf}), $\rho(s)\rightarrow 1$ & 0.045 & 0.084 & 0.095 & 0.242 & 0.106 & 0.059 & 0.025 \\
\hline %
\end{tabular}%
\end{table}%
\end{center}%
\vspace{-1cm}
In the last row of the previous table, we also give the outcome if $\rho(s)$ is set to 1, which would be the interaction-free limit of this function. We see that this would be a bad approximation if the interaction is enhanced as e.g. in the resonance region.\\
\newpage
We point out that the expression in Eq.~(\ref{eq:rhosol}) should be used instead of the original definition $\rho\sim4\left(\frac{2\pi}{L}\right)^2/\Delta s$ stemming from the ``brute force'' discretization of the $s'$-integral, because some information on the interaction in the infinite-volume limit (derived only from the spectrum) has already been implemented in (\ref{eq:rhosol}). We demonstrate this in the plot below.
\begin{center}
\begin{figure}[h]
\includegraphics[width=8.5cm]{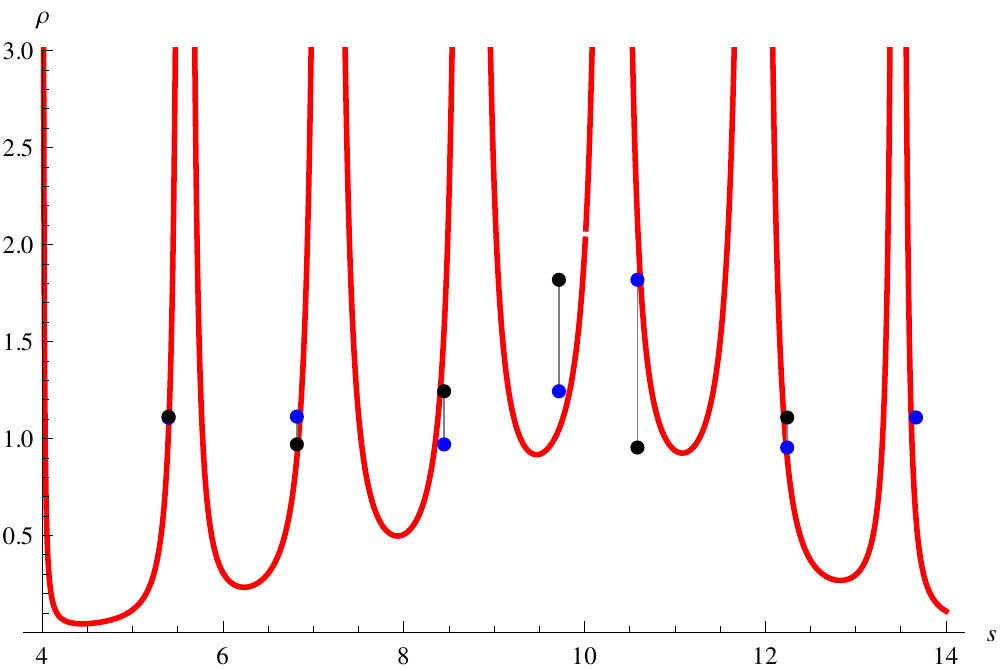}
\end{figure}
\end{center}%
\vspace{-1cm}
The red curve shows the function according to Eq.~(\ref{eq:rhosol}) (but note that it is {\em not\,} the graph of $\rho$, which is only defined at the $s_{j}$). The black and the blue points mark the values $\rho(s_{j})=4\left(\frac{2\pi}{L}\right)^2/(s_{j+1}-s_{j})$ and $\rho(s_{j})=4\left(\frac{2\pi}{L}\right)^2/(s_{j}-s_{j-1})$, respectively, which should lead to the same limit function $\rho(s)$. One observes that the uncertainty involved in the discretization of the $s'$-integral is quite large in the region where the phase shift rapidly varies from one energy eigenvalue to the next: there, the energy discretization cannot resolve the rapid variation of the final-state interaction in the infinite volume.\\
From our numerical results for $\delta_{0}(s_{j})$ and $|B(s_{j})|$, we can now infer the complex values of the form-factor $B(s_{j})$. These results are collected in the following tables.
\begin{center}
\vspace{-0.5cm}
\begin{table}[h]
\label{tab:Bresult1}
\begin{tabular}{|c||c|c|c|c|c|c|c|c|c|}
\hline
 $j$  & 1 & 2 & 3 & 4  \\
\hline
 $B(s_{j})$ (``num.'') & \,$2.806+1.054\,i$\, & \,$2.951+1.705\,i$\, & \,$3.048+3.343\,i$\, & \,$1.041+5.688\,i$\,  \\
\hline %
\end{tabular}%
\end{table}%
\end{center}%
\vspace{-1.5cm}
\begin{center}
\vspace{-0.5cm}
\begin{table}[h]
\label{tab:Bresult2}
\begin{tabular}{|c||c|c|c|c|c|c|c|}
\hline
 $j$  & 5 & 6 & 7   \\
\hline
 $B(s_{j})$ (``num.'') &\,$-2.091+4.894\,i$\, & \,$-2.264+1.102\,i$\, & \,$-1.271+0.289\,i$\, \\
\hline %
\end{tabular}%
\end{table}%
\end{center}%
\vspace{-1cm}
Now we have to use some parameterization of the form-factor in order to be able to perform an analytic continuation to the complex plane. We will again simply use our model amplitude (Eq.~(\ref{eq:fullM})), though in practice one will usually have to resort to some effective parameterizations, which might entail a considerable uncertainty. Fitting to the values in the above table (with our results for $\alpha_{fit},\,g_{fit},\,m_{0,fit}$) yields
\begin{displaymath}
b_{fit} = 0.969\,,\quad \tilde{b}_{fit} = 0.661\,,
\end{displaymath}
very close to our ``true'' values. The continuation of our model amplitude to the second Riemann sheet (where $\bar{I}_{\phi\phi}(s)\rightarrow \bar{I}_{\phi\phi}(s)-\frac{i\sigma(s)}{16\pi}\,$) yields (compare Eq.~(\ref{eq:resBoverressqrtT}))
\begin{displaymath}
\mathrm{Res}_{s_{\sigma}}B = -7.958-2.947\,i\qquad\Rightarrow\qquad -\frac{\mathrm{Res}_{s_{\sigma}}B}{\sqrt{-\mathrm{Res}_{s_{\sigma}}T}} = 0.609 +  0.103\,i\,,
\end{displaymath}
while the ``true'' value is found (continuing our model with the true parameters to the pole on the second sheet) to be 
\begin{displaymath}
\sqrt{Z_{\sigma}}\tilde{\beta}(s_{\sigma}) = (1.051 + 0.006\,i)\cdot(0.583 + 0.098\,i) = 0.612 + 0.106\,i\,.
\end{displaymath}
This is the value of the matrix element for the ``$\sigma$ decay constant'' associated with the operator $\mathcal{S}(x)$. Apparently, the same procedure can be applied for other operators with different matrix elements for $\phi\phi$ and $\sigma$ (with the same level density function $\rho(s)$, which depends only on the scattering phase and can thus be determined from the finite-volume spectrum alone). While the model discussed here is too simple to draw any general conclusions about realistic applications, we still believe that the main features of the physical problem in question can nicely be demonstrated in this framework. We hope that it has become clear that there is nothing mysterious to the relation between physical resonances and finite-volume lattice data in the resonant channel (or at least, it is not more mysterious than the concept of the complex energy plane). Additional problems appearing in realistic applications are the possibility of multiple open decay channels, possibly with more than two particles in the final states, discretization (lattice spacing) effects, and the use of smearing for source and sink operators. For the discussion of these more involved quaestions, we refer the reader to the literature cited in the introduction, and to future work along the lines of the present study.\\

\acknowledgments{I thank Maxim Mai, Andreas Sch\"afer and Philipp Wein for discussions on the manuscript. This work was supported by the Deutsche Forschungsgemeinschaft SFB/Transregio 55.}

\newpage
\begin{appendix}
\section{Some useful integrals}
\label{app:moreints}
\def\theequation{\Alph{section}.\arabic{equation}}
\setcounter{equation}{0}
For real $t\not=0$, $b>0$, we find the result
\begin{eqnarray}
\int_{b}^{\infty}dq\,\frac{\cos qt}{q}\sqrt{q^2-b^2} &=& \mathrm{sgn}(t)\frac{\pi}{2}b\left(J_{1}(bt)\left(bt\frac{\pi}{2}H_{0}(bt) -1\right) + bt J_{0}(bt)\left(1 -\frac{\pi}{2}H_{1}(bt)\right)\right)-\frac{\pi}{2}b \nonumber \\
 &=& \frac{\pi}{2}b\left(\frac{b|t|}{2}\,_{1}F_{2}\left(\frac{1}{2}\,;\,\frac{3}{2},2\,;\,-\frac{1}{4}(bt)^2\right)-1\right)\,.\label{eq:q0intt}
\end{eqnarray}
where $J_{n}$ and $H_{m}$ are the Bessel J and Struve H functions. Taking derivatives with respect to $t$, we obtain results for more integrals (which appear to be diverging on first sight),
\begin{eqnarray}
\int_{b}^{\infty}dq\,\sin\left(qt\right)\sqrt{q^2-b^2} &=& -b^2\frac{\pi}{2}\frac{J_{1}(bt)}{b|t|}\,,\label{eq:J1}\\
\int_{b}^{\infty}dq\,q\cos\left(qt\right)\sqrt{q^2-b^2} &=& -b^3\frac{\pi}{2}\frac{J_{2}(bt)}{b|t|}\,.\label{eq:J2}
\end{eqnarray}
For real $t$ (which we always assume here), we also find a result in terms of Bessel's Y,
\begin{equation}\label{eq:Y1}
\int_{b}^{\infty}dq\,\cos\left(qt\right)\sqrt{q^2-b^2} = b^2\frac{\pi}{2}\frac{Y_{1}(b|t|)}{b|t|}\,,
\end{equation}
and together with Eq.~(\ref{eq:J1})
\begin{equation}
\int_{b}^{\infty}dq\,e^{iqt}\sqrt{q^2-b^2} = b^2\frac{\pi}{2}\frac{(Y_{1}(b|t|)-i\,\mathrm{sgn}(t)J_{1}(b|t|))}{b|t|}\,.
\end{equation}
A generalization of Eq.~(\ref{eq:q0intt}) is
\begin{eqnarray}
h_{n}(t,b) &:=& \int_{b}^{\infty}dq\,\frac{\cos qt}{q^{2n+1}}\sqrt{q^2-b^2} =  |t|^{2n-1}\sin\left(n\pi\right)\Gamma(1-2n)\,_{1}F_{2}\left(-\frac{1}{2}\,;\,n,n+\frac{1}{2}\,;\,-\frac{1}{4}(bt)^2\right) \nonumber \\
 &+&  b^{1-2n}\frac{\sqrt{\pi}}{4}\frac{\Gamma\left(n-\frac{1}{2}\right)}{\Gamma(n+1)}\,_{1}F_{2}\left(-n\,;\,\frac{1}{2},\frac{3}{2}-n\,;\,-\frac{1}{4}(bt)^2\right)\,.\label{eq:hn}
\end{eqnarray}
Note that the limits $\lim\limits_{n \rightarrow 0}{h_{n}(t,b)}$, $\lim\limits_{n \rightarrow 1}{h_{n}(t,b)}$ etc. exist. For $n\rightarrow 0$, one recovers (\ref{eq:q0intt}). The asymptotic expansions of the Bessel and Struve functions are well-known, and we find that, for large $bt\gg1$, 
\begin{displaymath}
h_{0}(t,b)\,\rightarrow\,-\sqrt{\frac{\pi}{2b}}\frac{1}{t^{3/2}}\sin\left(bt+\frac{\pi}{4}\right)\,,
\end{displaymath}
and since $\frac{\partial^2}{\partial t^2}h_{n+1}(t,b)=-h_{n}(t,b)$, the asymptotic form of the $h_{n}$ must be in general
\begin{equation}\label{eq:hnasymp}
h_{n}(t,b)\,\rightarrow\,-\frac{1}{b^{2n}}\sqrt{\frac{\pi}{2b}}\frac{1}{t^{3/2}}\sin\left(bt+\frac{\pi}{4}\right)\,,
\end{equation}
neglecting all suppressed powers of $t^{-1}$. Note that there can not be an additional polynomial part because, for $t>0$, $h_{n}(t,b)$ is bounded by its value as $t\rightarrow 0$, 
\begin{displaymath}
|h_{n}(t,b)|< \lim\limits_{t \rightarrow 0}{h_{n}(t,b)}=b^{1-2n}\frac{\sqrt{\pi}}{4n!}\left|\Gamma\left(n-\frac{1}{2}\right)\right|\,.
\end{displaymath}
\newpage
We turn to a much more complicated integral,
\begin{equation}\label{eq:q0intt2}
g(t;a,b):=\int_{b}^{\infty}dq\,\frac{\sqrt{q^2-b^2}}{\sqrt{q^2-a^2}}\cos qt\,,\quad\mathrm{for}\quad b>0\,,\quad 0\leq a<b\,.
\end{equation}
It is obviously even in $t$ and $a$. The integral does not exist for $t=0$. However, taking the limit $t\rightarrow 0+$ along the real line, it can be checked that it tends to $g(t\rightarrow 0+;a,b)\rightarrow -bE\left(\frac{a^2}{b^2}\right)$, where $E(k^2)=\frac{\pi}{2}\,_{2}F_{1}(\frac{1}{2},-\frac{1}{2};1;k^2)$ is the well-known elliptic integral. The expression for $g(t;0,b)$ is given in (\ref{eq:q0intt}). Since $q^2\geq b^2>a^2$, we can expand the square-root in the denominator of the integrand of (\ref{eq:q0intt2}) and integrate term by term, using (\ref{eq:hn}):
\begin{equation*}
g(t;a,b) = \int_{b}^{\infty}dq\,\cos qt\,\sum_{n=0}^{\infty}\frac{\Gamma\left(n+\frac{1}{2}\right)}{\Gamma\left(\frac{1}{2}\right)\Gamma(n+1)}\frac{\sqrt{q^2-b^2}}{q^{2n+1}}a^{2n} 
 = \sum_{n=0}^{\infty}\frac{\Gamma\left(n+\frac{1}{2}\right)}{\Gamma\left(\frac{1}{2}\right)\Gamma(n+1)}h_{n}(t,b)a^{2n}\,.
\end{equation*}
According to Eq.~(\ref{eq:q0intt2}), a term linear in $|t|$ can only be generated from $h_{0}$ and $h_{1}$. Carefully taking the limits $n\rightarrow 0$ and $n\rightarrow 1$, one finds $g(t;a,b)=-bE(a^2/b^2) + \frac{\pi}{4}(b^2-a^2)|t|+\mathcal{O}(|t|^2)$. Reordering this series, it can also be shown that $\lim\limits_{a \rightarrow \pm b}{g(t;a,b)}=-\sin(bt)/t$ for $a\in\,]-b,b\,[$.\\
Using Eq.~(\ref{eq:hnasymp}), we deduce that the asymptotic form for large $t$ is given by
\begin{equation}\label{eq:gasymp}
g(t;a,b)\quad\overset{t\rightarrow\infty}{\longrightarrow}\quad -\sqrt{\frac{\pi}{2b}}\frac{b}{\sqrt{b^2-a^2}}\frac{\sin\left(bt+\frac{\pi}{4}\right)}{t^{3/2}}\,.
\end{equation}
Integrating Eq.~(\ref{eq:Y1}) over $t$, mathematica gives the result
\begin{equation}
\int_{b}^{\infty}dq\,\frac{\sin\left(qt\right)}{q}\sqrt{q^2-b^2} = \frac{\pi}{4}b\,G_{2,4}^{2,1}\left(\frac{bt}{2},\frac{1}{2}\biggr|\begin{array}{cccc} & 1,& -1 & \\ -\frac{1}{2}, & \frac{1}{2}; & -1, & 0 \end{array}\right)\,,
\end{equation}
in terms of the (generalized) Meijer-G function. More generally, similar to Eq.~(\ref{eq:hn})
\begin{eqnarray}
\tilde{h}_{n}(t,b) &:=& \int_{b}^{\infty}dq\,\frac{\sin qt}{q^{2n+1}}\sqrt{q^2-b^2} =  t^{2n-1}\cos\left(n\pi\right)\Gamma(1-2n)\,_{1}F_{2}\left(-\frac{1}{2}\,;\,n,n+\frac{1}{2}\,;\,-\frac{1}{4}(bt)^2\right) \nonumber \\
 &+&  b^{2-2n}t\frac{\sqrt{\pi}}{4}\frac{\Gamma\left(n-1\right)}{\Gamma(n+\frac{1}{2})}\,_{1}F_{2}\left(\frac{1}{2}-n\,;\,\frac{3}{2},2-n\,;\,-\frac{1}{4}(bt)^2\right)\,.\label{eq:hntilde}
\end{eqnarray}
Again, one can check that the limits $\lim\limits_{n \rightarrow 0}{\tilde{h}_{n}(t,b)}$, $\lim\limits_{n \rightarrow 1}{\tilde{h}_{n}(t,b)}$ etc. exist. From the known asymptotic behavior of the hypergeometric functions, we can infer that, for large $t$, 
\begin{equation}\label{eq:hntildeasymp}
h_{n}(t,b)\,\rightarrow\,\frac{1}{b^{2n}}\sqrt{\frac{\pi}{2b}}\frac{1}{t^{3/2}}\cos\left(bt+\frac{\pi}{4}\right)\,,
\end{equation}
and therefore, similar to Eq.~(\ref{eq:gasymp}),
\begin{equation}\label{eq:gtildeasymp}
\int_{b}^{\infty}dq\,\frac{\sqrt{q^2-b^2}}{\sqrt{q^2-a^2}}\sin qt\quad\overset{t\rightarrow\infty}{\longrightarrow}\quad \sqrt{\frac{\pi}{2b}}\frac{b}{\sqrt{b^2-a^2}}\frac{\cos\left(bt+\frac{\pi}{4}\right)}{t^{3/2}}\,.
\end{equation}
\newpage
Taking repeated time derivatives of the above results, one also shows
\begin{equation}\label{eq:expintasymp}
\int_{b}^{\infty}dq\,\frac{\sqrt{q^2-b^2}}{\sqrt{q^2-a^2}}\,q^{n}e^{-iqt}\quad\overset{t\rightarrow\infty}{\longrightarrow}\quad \sqrt{\frac{\pi}{2b}}\frac{b^{n+1}}{\sqrt{b^2-a^2}}\frac{e^{-ibt}}{(it)^{3/2}}\,,
\end{equation}
and by taking derivatives w.r.t. $a^2$,
\begin{equation}\label{eq:expdintasymp}
\int_{b}^{\infty}dq\,\frac{\sqrt{q^2-b^2}}{\sqrt{q^2-a^2}^3}\,q^{n}e^{-iqt}\quad\overset{t\rightarrow\infty}{\longrightarrow}\quad \sqrt{\frac{\pi}{2b}}\frac{b^{n+1}}{\sqrt{b^2-a^2}^3}\frac{e^{-ibt}}{(it)^{3/2}}\,,\quad\mathrm{etc.}
\end{equation}
\vspace{0.1cm}
In addition to the Fourier integrals discussed above, it is useful to know
\begin{equation}
\int_{0}^{\infty}dq\,\frac{\cos qt}{\left(q^2+M^2\right)^{n}} = \frac{\sqrt{\pi}\,t^{n-\frac{1}{2}}}{(2\sqrt{M^2})^{n-\frac{1}{2}}\Gamma(n)}K_{\frac{1}{2}-n}(\sqrt{M^2}\,t)\,,\quad n>0\,.
\end{equation}
The modified Bessel functions of half-integer degree can be expressed through exponential functions. For example, one has $K_{\frac{1}{2}}(z)=K_{-\frac{1}{2}}(z)=\sqrt{\pi}e^{-z}/\sqrt{2z}$ and thus, for example
\begin{equation}\label{eq:cosintM}
\int_{0}^{\infty}dq\,\frac{\cos qt}{q^2+M^2} = \frac{\pi}{2\sqrt{M^2}}e^{-\sqrt{M^2}|t|}\,,
\end{equation}
for real $M\not=0$. It is straightforward to see that such formulae can be analytically continued to imaginary $M$, at the cost of the introduction of an $i\epsilon$-prescription. For real $z$, we find from the theorem of residues
\begin{equation}\label{eq:cosintz}
\int_{0}^{\infty}dq\,\frac{\cos qt}{q^2-z^2\pm i\epsilon} = \mp i\pi \frac{e^{\mp i|zt|}}{2|z|}\,.
\end{equation}
Similarly, decomposing the integrand into partial fractions,
\begin{equation}\label{eq:cosintzz}
\int_{0}^{\infty}dq\,\frac{\cos qt}{(q^2-z_{1}^2 + i\epsilon)(q^2-z_{2}^2 + i\epsilon)} = \frac{i\pi}{z_{2}^{2}-z_{1}^{2}}\left(\frac{e^{-i|z_{1}t|}}{2|z_{1}|} - \frac{e^{-i|z_{2}t|}}{2|z_{2}|}\right)\,,\quad z_{1}\not= z_{2}\,,
\end{equation}
\begin{equation*}\label{eq:cosintzzsqr}
\int_{0}^{\infty}dq\,\frac{\cos qt}{(q^2-z_{1}^2 + i\epsilon)^2(q^2-z_{2}^2 + i\epsilon)} =  \frac{i\pi}{(z_{2}^{2}-z_{1}^{2})^2}\left((3z_{1}^{2}-z_{2}^{2}-i|z_{1}t|(z_{2}^{2}-z_{1}^{2}))\frac{e^{-i|z_{1}t|}}{4|z_{1}|^3} - \frac{e^{-i|z_{2}t|}}{2|z_{2}|}\right)\,.
\end{equation*}

\newpage

\section{A Fourier integral in $\mathbf{d=4}$}
\label{app:fourier}
\def\theequation{\Alph{section}.\arabic{equation}}
\setcounter{equation}{0}
Consider the integral
\begin{displaymath}
I_{M}(x-y) := \int\frac{d^4l}{(2\pi)^{4}}\frac{ie^{-il\cdot(x-y)}}{l^2-M^2+i\epsilon}\,.
\end{displaymath} 
The integral diverges if the space-time distance four-vector $x-y$ approaches zero. First, we choose a time-like distance here, and set $x-y=(t,\mathbf{0})$, with some $t>0$. We perform the $l^{0}$-integration by the method of residues, closing the contour in the lower complex $l^{0}$-plane:
\begin{eqnarray}
I_{M}(x-y) &=& \int\frac{d^{3}\mathbf{l}}{(2\pi)^3}\int_{-\infty}^{+\infty}\frac{dl^{0}}{2\pi}\frac{ie^{-il^{0}t}}{\left[l^{0}-\left(\sqrt{|\mathbf{l}|^2+M^2}-i\epsilon\right)\right]\left[l^{0}-\left(-\sqrt{|\mathbf{l}|^2+M^2}+i\epsilon\right)\right]} \nonumber \\
 &=& \frac{1}{4\pi^2}\int_{0}^{\infty}\frac{p^2dp}{\sqrt{p^2+M^2-i\epsilon}}e^{-i\sqrt{p^2+M^2}t} \label{eq:imformula}\\
 &=& \frac{M^2}{4\pi^2}\int_{0}^{\infty}d\phi\,\sinh^2\phi\,e^{-(i(M-i\epsilon)t)\cosh\phi} = \frac{M^2}{4\pi^2}\left(\frac{K_{1}(iMt)}{iMt}\right) = \frac{MK_{1}(iMt)}{4\pi^2it}\,.\nonumber
\end{eqnarray}
We have used the substitution $\frac{p}{M}=\sinh\phi$, and an integral formula\footnote{See e.g. \cite{WW:1996}, pp. 367, 377. We have corrected for an unusual sign convention factor $(-1)^{\nu}$ in the definition of the $K_{\nu}(z)$ used in that book, so that our signs agree with those used by mathematica\textsuperscript{{\textregistered}}.} for the modified Bessel functions of the second kind, $K_{\nu}(z)$, which obey
\begin{eqnarray}
\frac{2\nu}{z}K_{\nu}(z) &=& K_{\nu +1}(z)-K_{\nu -1}(z)\,,\label{eq:BesselKa}\\
\frac{dK_{\nu}(z)}{dz} &=& -\frac{1}{2}\left(K_{\nu -1}(z)+K_{\nu +1}(z)\right)\,.\label{eq:BesselKb}
\end{eqnarray}
Note that the result for $I_{M}(x-y)=I_{M}(t)$ is real for $t=-i\tau$ (euclidean time $\tau$), and real $M$. 
Let us also look at the general case of Eq.~(\ref{eq:imformula}), for $t:=x^{0}-y^{0}$, $r:=|\mathbf{x}-\mathbf{y}|$.
Again, we perform the $l^{0}$-integration by the method of residues, closing the contour in the lower complex $l^{0}$-plane (assuming $\mathrm{Re}\,t>0$). We also perform the angular integrations, to find
\begin{equation}
I_{M}(x-y) = \int_{0}^{\infty}\frac{|\mathbf{l}|^2d|\mathbf{l}|}{(2\pi)^3}\frac{2\pi}{2\sqrt{|\mathbf{l}|^2+M^2}}\left(\frac{e^{i|\mathbf{l}|r}-e^{-i|\mathbf{l}|r}}{i|\mathbf{l}|r}\right)e^{-it\sqrt{|\mathbf{l}|^2+M^2}}\,. \nonumber 
\end{equation}
We employ the substitutions $|\mathbf{l}|/M=\sinh\phi$, $t=s\cosh\xi$, $r=s\sinh\xi$ ($\Rightarrow s=\sqrt{t^2-r^2}$), and make use of the fact that the resulting integrand is even in $\phi$,
\begin{equation}
I_{M}(x-y) = \frac{M}{16\pi^2is}\int_{-\infty}^{+\infty}d\phi\,\frac{\sinh\phi}{\sinh\xi}\left(e^{iMs\sinh\phi\sinh\xi}-e^{-iMs\sinh\phi\sinh\xi}\right)e^{-iMs\cosh\phi\cosh\xi}\,.
\end{equation}
Noting that $\cosh\phi\cosh\xi\pm\sinh\phi\sinh\xi=\cosh\left(\phi\pm\xi\right)$, and substituting $\phi'=\phi\mp\xi$ in the first and second term, respectively, we find
\begin{eqnarray}
I_{M}(x-y) &=& \frac{M}{16\pi^2is}\int_{-\infty}^{+\infty}d\phi\,\frac{\sinh\phi}{\sinh\xi}\left(e^{-iMs\cosh\left(\phi-\xi\right)}-e^{-iMs\cosh\left(\phi+\xi\right)}\right) \\
 &=& \frac{M}{16\pi^2is}\int_{-\infty}^{+\infty}d\phi'\,\frac{\sinh\left(\phi'+\xi\right)-\sinh\left(\phi'-\xi\right)}{\sinh\xi}e^{-iMs\cosh\phi'} \nonumber \\
 &=& \frac{M}{8\pi^2is}\int_{-\infty}^{+\infty}d\phi'\,\cosh\phi'\,e^{-iMs\cosh\phi'} \overset{(\ref{eq:imformula})}{=} \frac{M}{8\pi^2is}2K_{1}\left(iMs\right) = \frac{K_{1}\left(iM\sqrt{t^2-r^2}\right)}{4\pi^2i\sqrt{t^2-r^2}}\,.\nonumber
\end{eqnarray}
We have again used the fact that the integrand is even under $\phi'\leftrightarrow -\phi'$, and partial integration using
\begin{displaymath}
\frac{d}{d\phi}\frac{1}{\cosh\phi} = -\frac{\sinh\phi}{\cosh^2\phi}\,,\quad \int_{0}^{\infty}d\phi\,z\sinh^2\phi\,e^{-z\cosh\phi} = \int_{0}^{\infty}d\phi\,\cosh\phi\,e^{-z\cosh\phi}=K_{1}(z)\,.
\end{displaymath}
This proves the general version of Eq.~(\ref{eq:imformula}), which could of course also be deduced from the special case above and Lorentz invariance of the integrand.

\end{appendix}

\end{document}